\documentclass[twocolumn,showpacs,preprintnumbers,amsmath,amssymb,prb,aps]{revtex4-1}
\usepackage{epsfig}
\usepackage{epstopdf}
\usepackage{graphicx}
\usepackage{dcolumn}
\usepackage{bm}
\usepackage{slashbox}
\usepackage{textcomp}
\usepackage{array}
\usepackage{subcaption}
\usepackage{booktabs}

\begin{document}

\title{Dependency of XC functionals and role of 3\textit{s}(2\textit{p}) orbitals of Co(Si) as core/valence states on the vibrational and thermodynamic properties of CoSi}
\author{Shamim Sk$^{1,}$}
\altaffiliation{Electronic mail: shamimsk20@gmail.com}
\author{Sudhir K. Pandey$^{2,}$}
\altaffiliation{Electronic mail: sudhir@iitmandi.ac.in}
\affiliation{$^{1}$School of Basic Sciences, Indian Institute of Technology Mandi, Kamand - 175075, India}
\affiliation{$^{2}$School of Engineering, Indian Institute of Technology Mandi, Kamand - 175075, India}


\begin{abstract}
First-principles phonon calculations along with density functional theory (DFT) play an important role to study the dynamical and thermal properties of materials. Here, we investigate the effect of exchange correlation (XC) functionals on the vibrational and thermodynamic properties of CoSi. The role of 3\textit{s}(2\textit{p}) orbitals of Co(Si) as core/valence states on the phonon properties of this compound is also studied. Phonon calculations are carried out by finite displacement method with supercell approach using equilibrium crystal structures obtained from DFT calculations. The calculated results are compared with the existing experiment. Three XC functionals, \textit{viz.}, LDA, PBEsol and SCAN are used for calculating the phonon dispersion, phonon density of states (DOS)/partial DOS and thermal properties of this compound. SCAN is found to give the highest phonon frequency of $\sim$56 meV which is in good agreement with the experimental value, while LDA (PBEsol) gives $\sim$54 ($\sim$55) meV. The zero-point energy is calculated as $\sim$ 10 kJ/mol for all the functionals. The Debye temperatures ($\Theta_{D}$) are computed as $\sim$626 K, $\sim$638 K and $\sim$650 K for LDA, PBEsol and SCAN, respectively. The $\Theta_{D}$ obtained from LDA gives the good agreement with the reported value. The phonon dispersion and phonon DOS are found to be dependent whether 3\textit{s}(2\textit{p}) orbitals of Co(Si) are considered as core or valence states. But, this orbital dependency is seemed to be insignificant on the thermal properties of this compound.

\vspace{0.3cm}

\end{abstract}

\maketitle
\section{INTRODUCTION}
The density functional theory (DFT)\cite{dft} is widely used computational scheme to study the electronic structure and related ground-state properties of materials. In this theory, the Kohn-Sham equation is solved to obtain the total ground-state energy. Kohn-Sham equation takes care the exchange and correlation term also to make the effective potential more realistic. The most accurate method for solving the Kohn-Sham equation is full potential linearized augmented plane wave (FP-LAPW). In this method, the basis set is made up of two parts: (1) inside the atomic sphere, where spherical potential is considered, and (2) interstitial region, where plane wave potential is employed. The most important term in Kohn-Sham equation is exchange correlation (XC) energy which is expressed as a functional of density, $E_{XC}[\rho(\textbf{r})]$. This term is approximated by different XC functionals. Among the large number of XC functionals developed, the local density approximation (LDA) and generalized gradient approximations (GGAs) are the popular functionals in condensed matter physics. The overall accuracy of the DFT calculation depends on how smartly functional has been chosen. 

In addition to the total ground-state energy, the DFT calculation also gives the total forces on each atom. By minimizing these total forces one can obtain the equilibrium crystal structure. This equilibrium crystal structure is used to generate the phonon frequency by displacing the atom from its equilibrium position. This type of method to analysis the phonon frequency is called finite displacement method (FDM)\cite{fdm}. Another method for phonon calculations is the density functional perturbation theory (DFPT)\cite{dfpt}. 

CoSi with a B20 simple cubic structure has been reported as a prominent thermoelectric (TE) candidate from last few decades\cite{asanable,kim,li,pan,skoug,sun}. The efficiency of TE materials can be evaluated by dimensionless parameter called \textit{figure-of-merit}, ZT. ZT is proportional to electrical conductivity ($\sigma$) and inversely proportional to the thermal conductivity ($\kappa$)\cite{zt}. For maximizing the efficiency, $\sigma$ should be increased with the simultaneous decrement of $\kappa$ which is quite diffucult to attain\cite{ashcroft,shamim}. One should keep in mind that once we increase the $\sigma$, the electronic part of the thermal conductivity ($\kappa_{el}$) will also increase according to the Wiedeman Franz law ($\kappa_{el} = L\sigma T$, where $L$ is Lorentz number.) Therefore, the smart way to maximize ZT is to optimize the lattice thermal conductivity ($\kappa_{l}$), which has an important contribution to total $\kappa$\cite{mahan}. Hence, it is necessary to study the phonon vibrations in order to optimize $\kappa_{l}$.    

The value of ZT for CoSi is moderate due to its high thermal conductivity as compared to other commercial thermoelectric materials like Bi$_{2}$Te$_{3}$,\cite{bite} PbTe\cite{pbte}. Therefore, reducing the $\kappa_{l}$ of CoSi is the signifying way to maximize ZT. For this, the study of lattice dynamics of this compound is strictly required. In this regard, the real test of XC functionals can be done by performing the phonon calculations using different functionals which is still missing from the literatures. Apart from this, the role of 3\textit{s}(2\textit{p}) orbitals of Co(Si) as core as well as valence states on the phonon calculations of CoSi using different XC functionals has also been investigated through the present study. The elastic and thermodynamic properties of this compound are studied by Liu \textit{et al}.\cite{liu} They have obtained the thermodynamic properties using quasi-harmonic Debye model. Povzner \textit{et al.}\cite{povzner} have developed a self-consistent thermodynamic model to study the thermal properties of CoSi. They\cite{povzner} have considered the influence of phonon anharmonicity in their calculations which was not taken into consideration in the work of Liu \textit{et al}\cite{liu}.

In this work, we have studied the phonon dispersion, phonon density of states (DOS)/Partial DOS (PDOS), phonon thermal properties of CoSi using FDM within the harmonic approximation. The role of 3\textit{s}(2\textit{p}) orbitals of Co(Si) as core as well as valence states are also investigated on the phonon calculations of this compound. Three XC functionals \textit{viz.}, LDA, PBEsol and SCAN are employed for the calculations. The specific heat at constant volume ($C_{V}$) and Helmholtz free energy ($F$) are calculated in the temperature range $0-1400$ K. The Zero-point energy is calculated as $\sim$ 10 kJ/mol for the compound. The Debye temperatures ($\Theta_{D}$) are also computed which are found to be $\sim$626 K, $\sim$638 K and $\sim$650 K for LDA, PBEsol and SCAN, respectively.

\section{COMPUTATIONAL DETAILS}
The calculations in the present work are performed using two computational tools. First of all, the total forces on the atoms are calculated using full potential linearized augmented plane wave (FP-LAPW) implemented in WIEN2k code\cite{wien2k}. The three exchange correlation (XC) functionals \textit{viz.}, LDA,\cite{lda} PBEsol\cite{pbesol} and SCAN\cite{scan} are employed for the calculations. The lattice parameters are taken from the literature\cite{boren}. The muffin-tin sphere radii for Co and Si are set to be 2.18 and 1.94 Bohr, respectively. To relax the compound, minimization of internal parameters have been done in k-mesh grid of size 8$\times$8$\times$8. During the relaxation, a force convergence criteria is fixed to be 1 mRy/Bohr. To calculate the forces on each atom for phonon calculations in 2$\times$2$\times$2 supercell, a k-mesh grid of size 4$\times$4$\times$4 is used. For this, the self-consistency is reached by giving the convergence of the total forces on atoms to be smaller than 0.1mRy/Bohr. All the parameters mentioned above are used for both the cases where 3\textit{s}(2\textit{p}) orbitals of Co(Si) are treated as a core or valence. 

The phonon dispersion, phonon DOS/PDOS and phonon thermal properties are calculated using phonopy package.\cite{phonopy} Phonon calculations are carried out by FDM\cite{fdm} using the equilibrium crystal structures obtained from DFT\cite{dft} calculations. In this work, the equilibrium crystal structures are obtained by minimizing the total forces on each atom using WIEN2k code\cite{wien2k}. For the analysis of phonon frequencies, here, we used supercell approaches along with FDM which are implemented in phonopy\cite{phonopy}. In this method each atom is displaced by 0.02 Bohr in x-direction which provides forces and hence a series of phonon frequencies. The calculations are executed by constructing of 2$\times$2$\times$2 supercell with 64 atoms in 21$\times$21$\times$21 k-mesh sampling. CoSi has a cubic B20 crystal structure with space group of P2$_{1}$3 (No. 198). The Co atom sits at Wyckoff position 4a (0.140, 0.140, 0.140), whereas Si atom occupies the Wyckoff position 4a (0.843, 0.843, 0.843)\cite{boren}. 

\section{RESULTS AND DISCUSSION}
In this entire section, we discuss the phonon dispersions, phonon DOS/PDOS and phonon thermal properties computed using three XC functionals \textit{viz.}, LDA, PBEsol and SCAN. The role of 3\textit{s}(2\textit{p}) orbitals of Co(Si) as core as well as valence states are also studied on the phonon calculations of CoSi. For the sake of simplicity, "3\textit{s}(2\textit{p})-core" and "3\textit{s}(2\textit{p})-valence" are named for which 3\textit{s}(2\textit{p}) orbitals of Co(Si) are considered as core and valence, respectively, in the whole section. 


Here, it is important to note that the experimental structure\cite{boren} were relaxed using all the functionals in WIEN2k code\cite{wien2k}. The coordinates of Co and Si of relaxed structures are tabulated in Table 1 along with the experimental coordinates for both the cases of 3\textit{s}(2\textit{p})-core and 3\textit{s}(2\textit{p})-valence. From the table, it is clear that there are no remarkable difference in the coordinates between the experiment and relaxed one. This small change in coordinates is not expected to affect the related properties significantly once we employ these functionals.  

\subsection{Phonon dispersion and phonon density of states}
Fig. 1 shows the computed phonon dispersions using three XC functionals \textit{viz.}, LDA, PBEsol and SCAN along the high-symmetry directions $\Gamma$-$X$-$M$-$\Gamma$-$R$-$M$-$X$ in the first Brillouin zone. Fig. 1(a)-(c) exhibit the phonon dispersions corresponding to 3\textit{s}(2\textit{p})-core, while Fig. 1(d)-(f) present for 3\textit{s}(2\textit{p})-valence. For any mechanically stable system the phonon frequencies (energies) should be real and positive under the harmonic approximation.\cite{phonopy} All the plots in Fig. 1 are showing the positive phonon energies suggesting that the compound is dynamically stable. As expected for this compound with eight atoms in the primitive cell, the phonon dispersion curve exhibits twenty four branches as shown in figures. Among them three are acoustic and remaining twenty one are optical branches. The acoustic branches along $\Gamma$-$M$ and $\Gamma$-$R$ directions are almost linear which signify that the group velocities are same as phase velocities in this region.\cite{ashcroft} 

\begin{table}
\caption{\small{The ionic coordinates of CoSi obtained from relaxed structures and from experiment. 3\textit{s}(2\textit{p})-core and 3\textit{s}(2\textit{p})-valence are labelled for which 3\textit{s}(2\textit{p}) orbitals of Co(Si) are considered as core and valence, respectively.}}
\resizebox{0.4\textwidth}{!}{%
\begin{tabular}{@{\extracolsep{\fill}}l c c c c c c c c c c c c c c} 
\hline\hline
 
\multicolumn{1}{c}{Methods} & & &  \multicolumn{4}{c}{3\textit{s}(2\textit{p})-core}  & & &   \multicolumn{4}{c}{3\textit{s}(2\textit{p})-valence} & \multicolumn{1}{c}{}\\ 
\cline{4-7} 
\cline{10-14}                                
  & & & \multicolumn{1}{c}{X$_{Co}$} & & & \multicolumn{1}{c}{X$_{Si}$}   & & &   \multicolumn{1}{c}{X$_{Co}$} & & & \multicolumn{1}{c}{X$_{Si}$}\\
   
\hline
LDA        &&& 0.1436 &&& 0.8423 &&& 0.1436 &&& 0.8423 \\
PBEsol     &&& 0.1445 &&& 0.8430 &&& 0.1448 &&& 0.8430 \\
SCAN       &&& 0.1454 &&& 0.8433 &&& 0.1454 &&& 0.8433 \\
Experiment\cite{boren} &&& 0.140 &&& 0.843 &&& 0.140 &&& 0.843 \\ [0.5ex]
 
\hline\hline
 
\end{tabular}}
\end{table} 

\begin{figure*}
\includegraphics[width=0.95\linewidth, height=8.0cm]{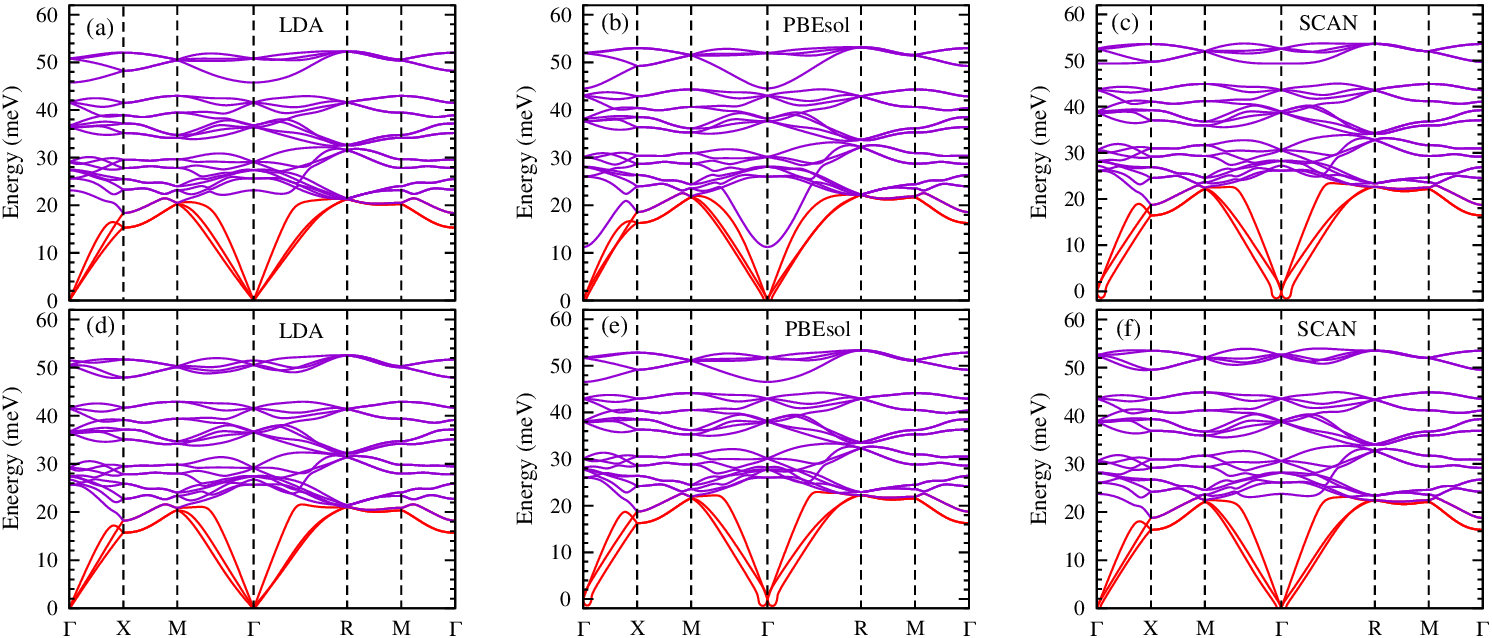} 
\caption{\small{Phonon dispersion for CoSi: 3\textit{s}(2\textit{p}) orbitals of Co(Si) are considered as (a)-(c) core and (d)-(f) valence.}}
\end{figure*} 

First of all, we discuss the dependency of XC functionals on the phonon dispersions. In the case of 3\textit{s}(2\textit{p})-core (Fig. 1(a)-(c)), features of one of the optical branches between energy range $\sim$ 45-50 meV are different for all the three functionals around $\Gamma$ point. This branch occurs at highest energy ($\sim$ 49 meV) for SCAN, whereas it appears at lowest energy ($\sim$ 45 meV) for PBEsol at $\Gamma$ point. At the same $\Gamma$ point, another one of the optical branches occurs at $\sim$ 11 meV for PBEsol, while this branch arises above $\sim$ 20 meV for both of LDA and SCAN. Few optical branches are nearly degenerate at $\sim$ 22 meV for LDA and PBEsol, whereas a finite separation between the branches is clearly observed for SCAN at $R$ point. In the case of 3\textit{s}(2\textit{p})-valence (Fig. 1(d)-(f)), all the optical branches for LDA and PBEsol arise above $\sim$ 25 meV, whereas one of the optical branches appears at $\sim$ 24 meV for SCAN at $\Gamma$ point. At the same point, a optical branch appears at $\sim$ 47 meV for PBEsol, while this branch occurs above $\sim$ 50 meV for LDA and SCAN. Therefore, the above discussion suggests that these three functionals are giving different phonon dispersions for both the cases of 3\textit{s}(2\textit{p})-core and 3\textit{s}(2\textit{p})-valence. Hence, the different phonon related properties of CoSi is expected once we use these functionals individually.

Now, we compare between the phonon dispersions for the cases of 3\textit{s}(2\textit{p})-core and 3\textit{s}(2\textit{p})-valence when using same functional. In Fig 1(a), two optical branches appear at $\sim$ 23 and $\sim$ 45 meV for 3\textit{s}(2\textit{p})-core, while these branches occur above $\sim$ 25 and $\sim$ 50 meV (Fig. 1(d)), respectively for 3\textit{s}(2\textit{p})-valence at $\Gamma$ point. In Fig. 1(b), the lowest energy optical branch arises at $\sim$ 11 meV for 3\textit{s}(2\textit{p})-core, whereas this branch occurs at $\sim$ 25 meV (Fig. 1(e)) in case of 3\textit{s}(2\textit{p})-valence at $\Gamma$ point. From Fig. 1(c) and (f), it is clear that the SCAN also gives the different features of dispersions for 3\textit{s}(2\textit{p})-core and 3\textit{s}(2\textit{p})-valence. From the above discussions, it is clear that the phonon dispersion of CoSi is dependent on XC functionals as well as 3\textit{s}(2\textit{p})-core and 3\textit{s}(2\textit{p})-valence cases. Therefore, experimental verification is required to compare our calculated phonon dispersions.

\begin{figure*}
\includegraphics[width=0.95\linewidth, height=6.5cm]{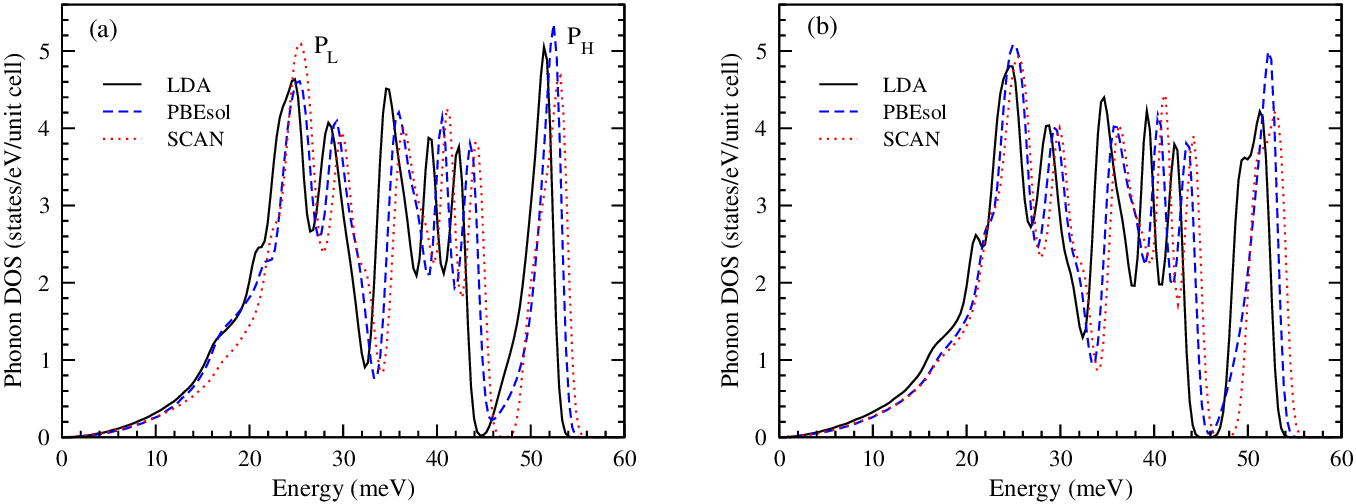} 
\caption{\small{Phonon density of states (DOS) for CoSi: 3\textit{s}(2\textit{p}) orbitals of Co(Si) are considered as (a) core and (b) valence. P$_{L}$ and P$_{H}$ denote for lowest energy peak and highest energy peak, respectively.}}
\end{figure*}

\begin{figure*}
\includegraphics[width=0.95\linewidth, height=8.0cm]{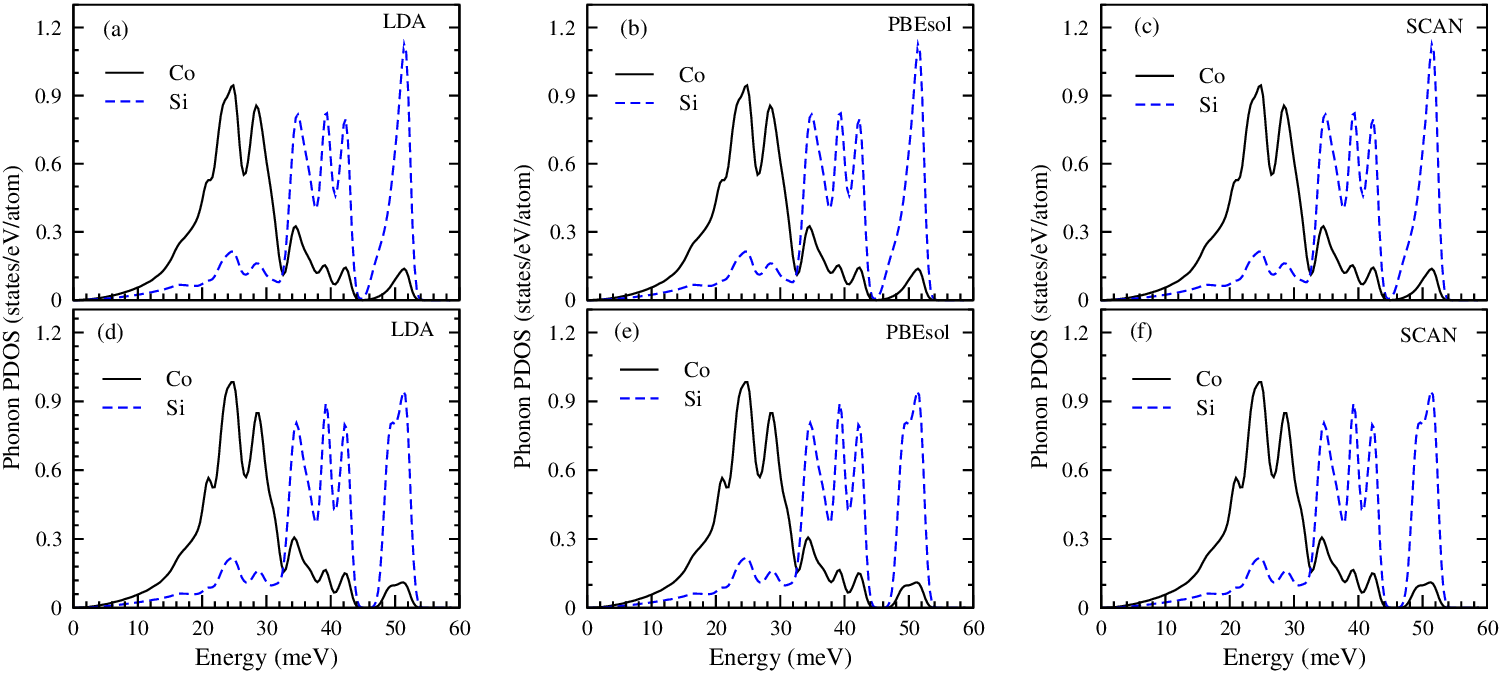} 
\caption{\small{Phonon partial density of states (PDOS) for CoSi: 3\textit{s}(2\textit{p}) orbitals of Co(Si) are considered as (a)-(c) core and (d)-(f) valence.}}
\end{figure*}

\begin{figure*}
\includegraphics[width=0.95\linewidth, height=6.8cm]{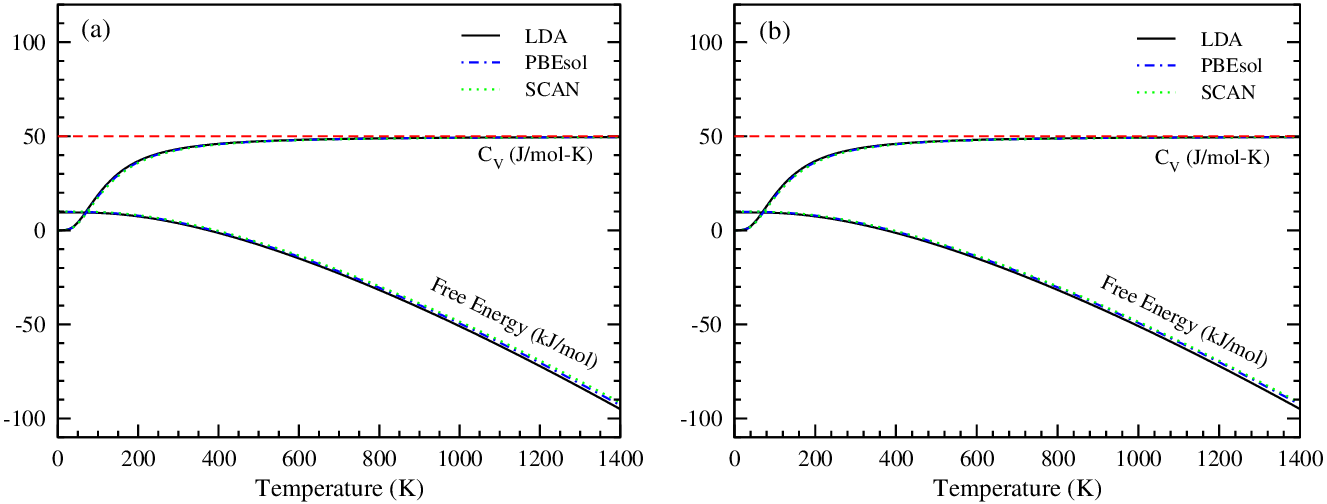} 
\caption{\small{Specific heat at constant volume (C$_{V}$) and Helmholtz free energy (F): 3\textit{s}(2\textit{p}) orbitals of Co(Si) are considered as (a) core and (b) valence.}}
\end{figure*}

Fig. 2 shows the calculated phonon density of states (DOS) for the cases of 3\textit{s}(2\textit{p})-core and 3\textit{s}(2\textit{p})-valence using all the three functionals. Fig. 2(a) displays phonon DOS corresponding to 3\textit{s}(2\textit{p})-core and Fig. 2(b) exhibits the phonon DOS of 3\textit{s}(2\textit{p})-core calculated using LDA, PBEsol and SCAN. From the figures, it is clear that all the peak positions are slightly deviated to each other for different functionals. Corresponding to crossing point of acoustic/optical branches, a more number of DOS peaks appear in the figures. The zero DOS at $\sim$ 45 meV signifies the separation between higher frequencies optical branches to the lower frequencies branches. In the energy region $\sim$ 45-54 meV, the highest energy DOS peak is appeared which corresponds to optical branches. These branches are expected to build up from the contribution of lighter Si atom mainly. In the low lying energy region ($\sim$0-20 meV), the amount of DOS for LDA and PBEsol are same for 3\textit{s}(2\textit{p})-core, while in the case of 3\textit{s}(2\textit{p})-valence, the amount of DOS for PBEsol and SCAN are same. In Fig. 2(b), at $\sim$ 50 meV a hump is observed, whereas this is absent in Fig. 2(a). Delaire \textit{et al.}\cite{delaire} have reported the phonon DOS obtained by inelastic neutron scattering at different temperature. We compare our phonon energy window, lowest (P$_{L}$) and highest (P$_{L}$) energy peaks with their work (10 K data) as shown in Table 2. From table, it is clear that energy window of SCAN is in good agreement with the experimental value. The lowest energy peak for all the functionals occur at slight higher energy than experiment. 

\begin{table}
\caption{\small{Comparison of phonon energy window, lowest (P$_{L}$) and highest (P$_{H}$) energy DOS peaks between calculated and experimental one. P$_{L}$ and P$_{H}$ are marked in Fig. 2(a).}}
\resizebox{0.4\textwidth}{!}{%
\begin{tabular}{@{\extracolsep{\fill}}l c c c c c c c c c c c c c c c} 
\hline\hline
 
\multicolumn{1}{c}{Methods} &&& \multicolumn{1}{c}{Energy window} &&& \multicolumn{4}{c}{P$_{L}$} &&& \multicolumn{4}{c}{P$_{H}$} \\                                 
  &&& \multicolumn{1}{c}{(meV)} &&&&& \multicolumn{1}{c}{(meV)}   &&&&&&&   \multicolumn{1}{c}{(meV)} \\
   
\hline
LDA        &&& 0-54 &&&&& 24.7 &&&&&&& 51.4  \\
PBEsol     &&& 0-55 &&&&& 25.3 &&&&&&& 52.4  \\
SCAN       &&& 0-56 &&&&& 25.3 &&&&&&& 53.0  \\
Experiment\cite{delaire} &&& 0-56 &&&&& 23 &&&&&&& 53.0  \\ [0.5ex]
 
\hline\hline
 
\end{tabular}}
\end{table}                       

To see the contribution to the phonon DOS from the different atoms, partial density of states (PDOS) are calculated as shown in Fig. 3. Fig. 3(a)-(c) exhibit the PDOS for 3\textit{s}(2\textit{p})-core and Fig. 3(d)-(f) show the PDOS for 3\textit{s}(2\textit{p})-valence using all the three functionals. In the energy region $\sim$ 0-32 meV, the contribution in DOS comes from heavier Co atom mainly with negligibly small contribution from Si atom. In this region the contribution in DOS from Co is $\sim$82$\%$, while for that of Si is $\sim$18$\%$. In the energy range $\sim$ 32-56 meV, the dominating contribution in DOS comes from lighter Si atom mainly. The contribution in DOS in this region from Co and Si are $\sim$18$\%$ and $\sim$82$\%$, respectively. In both the cases whether 3\textit{s}(2\textit{p}) is considered as core or valence, the heavier Co atom contributes to the DOS in lower energy region, whereas lighter Si atom contributes to the DOS in higher energy region mainly.

From the above discussion it is clear that the phonon dispersion and phonon DOS are dependent on XC functionals as well as 3\textit{s}(2\textit{p}) orbitals of Co(Si) whether these orbitals are considered as core or valence. Therefore, care should be taken in studying these properties of the compound.                

\subsection{Thermal properties}
Temperature dependent specific heat at constant volume ($C_{V}$) and Helmholtz free energy ($F$) are calculated using thermodynamic relations as follows\cite{dove},

\begin{equation}
C_{V}={\sum\limits_{\textbf{q}j}}k_{B}\bigg(\dfrac{\hbar\omega_{\textbf{q}j}}{k_{B}T}\bigg)^{2}\frac{\exp(\hbar\omega_{\textbf{q}j}/k_{B}T)}{[\exp(\hbar\omega_{\textbf{q}j}/k_{B}T)-1]^{2}},
\end{equation} 
and
\begin{equation}
F=\dfrac{1}{2}{\sum\limits_{\textbf{q}j}}\hbar\omega_{\textbf{q}j}+k_{B}T{\sum\limits_{\textbf{q}j}}\ln[1-\exp(-\hbar\omega_{\textbf{q}j}/k_{B}T)].
\end{equation} 

Where $k_{B}$, $\hbar$ and $T$ are the Boltzmann constant, the reduced Planck's constant and the temperature, respectively. $\omega_{\textbf{q}j}$ is the phonon frequency of mode \{$\textbf{q}$, $j$\}, where $\textbf{q}$ is wave vector and $j$ is band index. Eqns. 1 and 2 are used to compute the $C_{V}$ and $F$, respectively, which are implemented in phonopy code.\cite{phonopy} This method of calculating $C_{V}$ and $F$ are also executed by other works\cite{shastri,antik}. The calculated values of $C_{V}$ and $F$ in the temperature range $0-1400$ K are shown in Fig. 4(a) and (b) for 3\textit{s}(2\textit{p})-core and 3\textit{s}(2\textit{p})-valence, respectively. The dashed line in the figure denotes the classical Dulong and Petit's limit of $C_{V}$. For CoSi, this value is 50 J/mol-K. For both the cases of 3\textit{s}(2\textit{p})-core and 3\textit{s}(2\textit{p})-valence, Dulong and Petit's limit reaches at $\sim$ 700 K. The $F$ are also calculated as shown in the same Fig. 4. Quantum mechanics tells that in solids vibrational energy presents even at zero temperature which is called zero-point energy.\cite{hao} This energy can be obtained from the y-intercept of $F$ at 0 K.\cite{ashcroft,phonopy} Zero-point energy is found to be same for both the cases of 3\textit{s}(2\textit{p})-core and 3\textit{s}(2\textit{p})-valence with the corresponding value of $\sim$ 10 kJ/mol for all the functionals. From the figures, it is clear that $F$ is decreasing monotonically with the increasing of temperature in the full temperature range. The values of $F$ are found to be 0 at $\sim$374 K, $\sim$386 K and $\sim$392 K for LDA, PBEsol and SCAN, respectively for both the cases of 3\textit{s}(2\textit{p})-core and 3\textit{s}(2\textit{p})-valence. The value of $F$ at 1400 K are calculated as $\sim-$95, $\sim-$93 and $\sim-$92 kJ/mol for three functionals, respectively. 

The Debye temperature ($\Theta_{D}$) for this compound is also calculated. $\Theta_{D}$ is defined as the transition temperature at which all the phonon modes begin to excite.\cite{ashcroft} In this study, $\Theta_{D}$ is calculated using $\Theta_{D}$=$\hbar\omega_{D}/k_{B}$, where the maximum phonon frequency is considered as Debye frequency ($\omega_{D}$). Here, it is important to note that the value of maximum phonon frequency are same for both the cases of 3\textit{s}(2\textit{p})-core and 3\textit{s}(2\textit{p})-valence when using the same functional. From the phonon DOS of Fig. 2, the maximum phonon frequency ($\nu_{D}$) are found to be $\sim$54 meV, $\sim$55 meV and $\sim$56 meV for LDA, PBEsol and SCAN, respectively. Therefore, the corresponding $\Theta_{D}$ values are calculated as $\sim$626 K, $\sim$638 K and $\sim$650 K, respectively. The result of LDA agrees very well with the calculated value of $\Theta_{D}$ (= 625 K) by Petrova \textit{et al.}\cite{petrova} at very low temperature($T \rightarrow 0$).

Above discussion suggests that the effect of XC functionals on $F$ and $\Theta_{D}$ are come into account. Hence, attention should be paid in choosing the XC functionals for investigating these properties. Here, it also should be noted that the thermal properties are insignificant whether 3\textit{s}(2\textit{p}) orbitals of Co(Si) are considered as core or valence states.

\section{CONCLUSIONS}
In conclusion, the vibrational and thermodynamic properties of CoSi are studied using first-principles phonon calculations along with density funtional theory. Three exchange correlation functionals viz., LDA, PBEsol and SCAN are employed. Phonon calculations are done using finite displacement method with supercell approach within harmonic approximation. From the phonon dispersion and phonon DOS calculations, it is observed that SCAN gives the highest phonon frequency of $\sim$56 meV with good agreement of the experimental value, whereas LDA(PBEsol) gives $\sim$54($\sim$55) meV. The specific heat at constant volume (C$_{V}$) and Helmholtz free energy ($F$) are calculated in the temperature region $0-1400$ K. From temperature dependent of C$_{V}$, it is observed that the Dulong and Petit's limit of C$_{V}$ is reached at $\sim$ 700 K. The value of zero-point energy is computed as $\sim$ 10 kJ/mol for all the functionals. The Debye temperatures ($\Theta_{D}$) are also calculated which are found to be $\sim$626 K, $\sim$638 K and $\sim$650 K for LDA, PBEsol and SCAN, respectively. Therefore, care should be taken in choosing the XC functionals for studying the vibrational and thermodynamic properties of this compound. The significant effect on the vibrational properties of this compound is observed whether 3\textit{s}(2\textit{p}) orbitals of Co(Si) are considered as core or valence states. But, thermal properties of this compound is appeared to be insensitive whether 3\textit{s}(2\textit{p}) orbitals of Co(Si) are considered as core or valence.


\end{document}